\documentclass[aps,physrev,reprint,superscriptaddress]{revtex4-2}

\usepackage{graphicx}
\usepackage{dcolumn}
\usepackage{bm}
\usepackage{textgreek}

\begin{document}

\title{\textbf{Robustness Optimization for Compact Free-electron Laser Driven by Laser Wakefield Accelerators} 
}

\author{Hai Jiang}
\affiliation{School of Optical-Electrical and Computer Engineering, University of Shanghai for Science and Technology, Shanghai 200093, People's Republic of China}
\affiliation{State Key Laboratory of Ultra-intense laser Science and Technology, Shanghai Institute of Optics and Fine Mechanics (SIOM), Chinese Academy of Sciences (CAS), Shanghai 201800, People's Republic of China}

\author{Ke Feng}
 \email{Contact author: fengke@siom.ac.cn}
\affiliation{State Key Laboratory of Ultra-intense laser Science and Technology, Shanghai Institute of Optics and Fine Mechanics (SIOM), Chinese Academy of Sciences (CAS), Shanghai 201800, People's Republic of China}
 
\author{Runshu Hu}
\affiliation{State Key Laboratory of Ultra-intense laser Science and Technology, Shanghai Institute of Optics and Fine Mechanics (SIOM), Chinese Academy of Sciences (CAS), Shanghai 201800, People's Republic of China}
\affiliation{Center of Materials Science and Optoelectronics Engineering, University of Chinese Academy of Sciences, Beijing 101408, People's Republic of China}

\author{Qiwen Zhan}
 \email{Contact author: qwzhan@usst.edu.cn}
\affiliation{School of Optical-Electrical and Computer Engineering, University of Shanghai for Science and Technology, Shanghai 200093, People's Republic of China}

\author{Wentao Wang}
 \email{Contact author: wwt1980@siom.ac.cn}
\affiliation{State Key Laboratory of Ultra-intense laser Science and Technology, Shanghai Institute of Optics and Fine Mechanics (SIOM), Chinese Academy of Sciences (CAS), Shanghai 201800, People's Republic of China}
 
\author{Ruxin Li}
\affiliation{State Key Laboratory of Ultra-intense laser Science and Technology, Shanghai Institute of Optics and Fine Mechanics (SIOM), Chinese Academy of Sciences (CAS), Shanghai 201800, People's Republic of China}
\affiliation{Center of Materials Science and Optoelectronics Engineering, University of Chinese Academy of Sciences, Beijing 101408, People's Republic of China}
\affiliation{School of Physical Science and Technology, ShanghaiTech University, Shanghai 201210, People's Republic of China}

% \date{\today}

\begin{abstract}
Despite the successful demonstration of compact free electron lasers (FELs) driven by laser wakefield accelerators (LWFAs), the inherent shot-to-shot fluctuations in LWFAs, including both laser and plasma instabilities, remain a primary obstacle to realizing LWFA-driven FELs with robust operation. Here, we present a conceptual design for LWFA-driven FELs with sufficient tolerance against shot-to-shot fluctuations using the Covariance Matrix Adaptation Evolution Strategy (CMA-ES). Start-to-end simulations demonstrated that this systematic optimization resulted in a significant improvement in the robustness of FELs. With the optimized configurations, the radiation energy can be maintained above 1 \textmu J at a wavelength of approximately 25 nm, even when accounting for twice the root-mean-square (RMS) ranges of these instabilities. This proposed scheme represents a substantial advancement in the development of compact LWFA-driven FEL systems, enabling robust operation and paving the way for the realization of reliable and widely accessible sources.
\end{abstract}
\maketitle

\section{Introduction}
Free electron lasers (FELs) are capable of generating high-brilliance and coherent radiation across a broad range of wavelengths, extending even into the hard X-ray regime \cite{madey1971, emma2010, ishikawa2012, kang2017, decking2020}. Their unique capabilities hold the potential to revolutionize various scientific disciplines by enabling unparalleled atomic resolutions on femto-to-attosecond timescales \cite{bostedt2016, burnett2020}. These FEL facilities typically rely on the radio-frequency (RF) accelerators with large scales and incurs highly costs. With unprecedented accelerating gradients reaching several hundreds of gigavolt per meter \cite{Tajima1979, nakajima1995}, several orders of magnitude higher than those of RF accelerators, laser wakefield accelerators (LWFAs) provide a promising alternative for driving compact FELs. The realization of such compact LWFA-driven FELs has been identified as one of the major challenges in this decade, as addressed in the European Plasma Research Accelerator with eXcellence In Applications (EuPRAXIA) \cite{Assmann2020}. Motivated by the vast potential of these more accessible and cost-effective systems, numerous innovative schemes and state-of-the-art technologies have been proposed, resulting in significant improvements in beam qualities of the accelerated \textit{e} beams from LWFAs \cite{Esarey2009, Gonsalves2011, Corde2013, Buck2013, Wang2016, Downer2018, Maier2020, Ke2021, ferran2022, gotzfried2020}. Recently, compact FELs driven by LWFAs have been experimentally demonstrated, with the self-amplified spontaneous emission (SASE) configuration at 27 nm and seeded configuration at 269 nm \cite{Wang2021,Labat2023}. However, the limited reliability and reproducibility of LWFAs present a fundamental barrier to realize compact and operationally robust FELs for practical application \cite{Huang2007,Galletti2024}.

In LWFAs, the interaction between high-intensity laser pulses and plasma constitutes an extraordinarily complex, nonlinear process. Consequently, even modest fluctuations in the drive laser or plasma parameters, inevitably result in substantial shot-to-shot fluctuations in \textit{e}-beam qualities \cite{Esarey2009}. Moreover, these fluctuations in beam qualities, including energy, energy spread and emittance, degrade the beam during transport through the beamline, ultimately hinder the gain and pose significant robustness challenges for LWFA-driven FELs \cite{Huang2007}. To address these challenges, it is imperative to systematically quantify the beam quality fluctuations induced by the instabilities of both laser and plasma. Additionally, substantial optimization of the beamline is essential to enhance its tolerance to these shot-to-shot fluctuations, thereby ensuring robustness of LWFA-driven FELs. For optimizing such complex nonlinear systems with high evaluation costs, black-box optimizers have emerged as a particularly suitable approach. Various optimizers, including Bayesian optimization, genetic algorithms, and evolutionary strategies like Covariance Matrix Adaptation Evolution Strategy (CMA-ES), have been demonstrated as effective tools in the field of laser-plasma physics and accelerators \cite{Duris2020, Jalas2021, Kirchen2021, Dolier2022, Irshad2023, Döpp2023}.

In this paper, we propose a conceptual design for a high-gain LWFA-driven FELs with enhanced robustness and reliability. Through detailed numerical simulations, the shot-to-shot fluctuations in beam quality have been quantitatively analyzed, which are attributable to instabilities from both laser and plasma, including laser energy jitter, focal position displacement induced by wavefront distortion, and shock front position instability in the plasma. To address these fluctuations, we employed the CMA-ES algorithm for beamline optimization, with the corresponding objective function of the FEL energy. Start-to-end simulations show that the optimized system can maintain FEL radiation in the saturated or high-gain regime with corresponding energy exceeding 1 \textmu J in the extreme ultraviolet (EUV) region, even operating within twice the root-mean-square (RMS) range of each parameter jitter. Quantitative tolerance analysis further demonstrates the scheme's robustness to up to 1 mrad of beam pointing jitter while maintaining radiation energy above 1 \textmu J. This energy corresponds to more than $1\times10^{11}$ photons per pulse in EUV regime, satisfying the photon flux requirements for applications such as coherent diffractive imaging \cite{Seibert2011, Jiang2010}. As a conceptual design with clear pathways for future experiment, the proposed scheme represents a significant step toward robust, compact LWFA-driven FELs and offers a promising route to reliable and widely accessible systems.

\section{LWFA SIMULATION}
\begin{figure}
    \centering
    \includegraphics[width=1\linewidth]{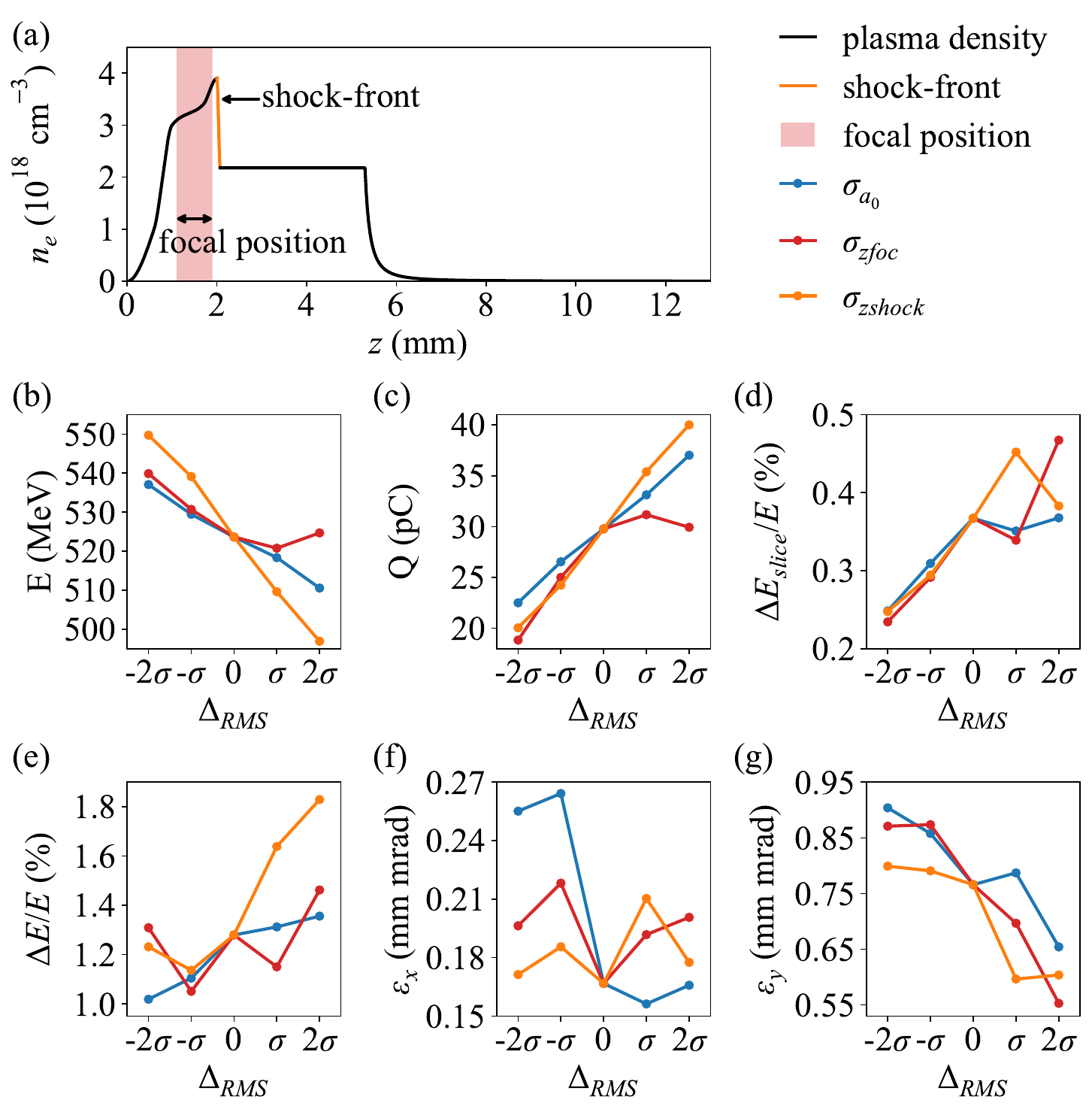}
    \caption{\label{fig:fig1}
    PIC simulation results of LWFA with inherent parameter jitters. (a) Schematic of plasma density profile and fluctuation sources, with the pink shaded area region and the orange solid line represent the range of focal position jitter and the initial shock-front position. (b)-(g) Electron beam properties with inherent parameter jitters: the blue, red and orange curves represent the laser energy variations, laser focal position displacement and shock front position instability, respectively. Horizontal axes denotes $\pm2$ folds RMS range of these three parameter jitters, and the vertical axes represent beam energy (b), charge (c), slice energy spread (d), global energy speread (e), horizontal (f) and vertical (g) normalized projected emittance.  
 }
\end{figure}

To investigate the shot-to-shot fluctuations in beam quality generated by LWFAs, quasi-three-dimensional (3D) particle-in-cell (PIC) simulations were conducted using the FBPIC code with a co-moving simulation box \cite{Lehe2016,Jalas2017}. The simulation domain has a size of 50 \textmu m and 120 \textmu m in the longitudinal and transverse directions, respectively, with the longitudinal grid size $\Delta z=31.25$ nm and the transverse grid size $\Delta r=80$ nm and 16 macroparticles per cell. In the simulation, we used a Gaussian laser pulse with a central wavelength of $\lambda=800$ nm, a waist radius of $\omega_0=35$ \textmu m, a pulse duration of 25 fs (full width at half maximum, FWHM), and a normalized vector potential of $a_{0}$ = 1.3. The plasma density profile employed in the simulation is shown in Fig.~\ref{fig:fig1}(a), with a peak shock density of $3.9\times10^{18}$ cm$^{-3}$, a plateau density of $2.2\times10^{18}$ cm$^{-3}$, and a shock width of 75 \textmu m \cite{Feng2025}.

The shot-to-shot fluctuations of the accelerated \textit{e} beams primarily originate from instabilities from both laser and plasma. These instabilities were quantitatively described laser energy jitter, focal position displacement induced by wavefront distortion, and shock front position instability in the plasma. In the presented simulation, the relative RMS laser energy jitter of $\Delta E/E=\pm0.54\%$ \cite{Wu2020} is applied, corresponding to the performance of the 200 TW Ti:sapphire laser system operating at 1 Hz \cite{Wu2020}. The focal position displacement jitter is set to an RMS value of $\sigma_{zfoc}=0.2$ mm, based on the reported values \cite{Maier2020}. For the shock front position, an RMS jitter of $\sigma_{zshock}=4.9$ \textmu m is adopted, following the experimental data provided in Appendix~\ref{app:A}. A comprehensive quantitative characterization of the \textit{e}-beam parameters under the influence of these instabilities of laser and plasma is presented in Figs.~\ref{fig:fig1}(b)-~\ref{fig:fig1}(g), including beam energy, charge, slice energy spread, global energy spread, as well as horizontal and vertical normalized projected emittance. It is noted that the laser energy jitter is represented by the variation of normalized vector potential $a_0$ in the following discussions. These jitters were varied up to ±2 times the RMS values in the simulations.  Each \textit{e}-beam parameter depicted in Figs.~\ref{fig:fig1}(b)-~\ref{fig:fig1}(g) containing 13 independent simulation runs, with $\Delta_{RMS}=0$ serving as the initial and jitter-free condition. 

By precisely controlling electron injection and chirp compensation through tailored plasma density profiles, high-quality electron beams can be obtained \cite{Feng2025}. However, these parameter jitters destabilizes the electron injection process, leading to substantial charge fluctuations. As shown in Fig.~\ref{fig:fig1}(c), when accounting for jitters in all these three parameters ($\sigma_{a_0}$, $\sigma_{zfoc}$, and $\sigma_{zshock}$), the relative charge fluctuation reaches $\pm22.0\%$ RMS. Such charge variations lead to distinct beamloading effects under different jitter conditions, which in turn cause changes in acceleration efficiency and inhomogeneities in the accelerating fields. Consequently, the beam quality degrades markedly, exhibiting the following RMS fluctuations: $\pm2.7\%$ in energy, $\pm21.2\%$ in slice energy spread, $\pm17.3\%$ in global energy spread, $\pm16.2\%$ and $\pm15.1\%$ in horizontal and vertical projected emittance, respectively. 

These instabilities fundamentally limit the radiation gain and operational stability of FELs, for which high-gain lasing generally demands three key beam properties: elevated charge, reduced slice energy spread, and minimized transverse emittance \cite{Huang2007}. Our analysis reveals an intrinsic trade-off among these three metrics, as shows by their anti-correlated trends in Figs.~\ref{fig:fig1}(c), ~\ref{fig:fig1}(d) and ~\ref{fig:fig1}(g). Strategic design and optimization of the beamline layout can effectively mitigate beam quality discrepancies, thereby defining a workable parameter space that supports stable, high-gain FEL radiation. This approach compensates for the inherent fluctuations in electron beam characteristics, directly addressing a critical challenge in achieving consistent FEL performance.

\section{BEAMLINE OPTIMIZATION}
\begin{figure*}
    \centering
    \includegraphics[width=1\linewidth]{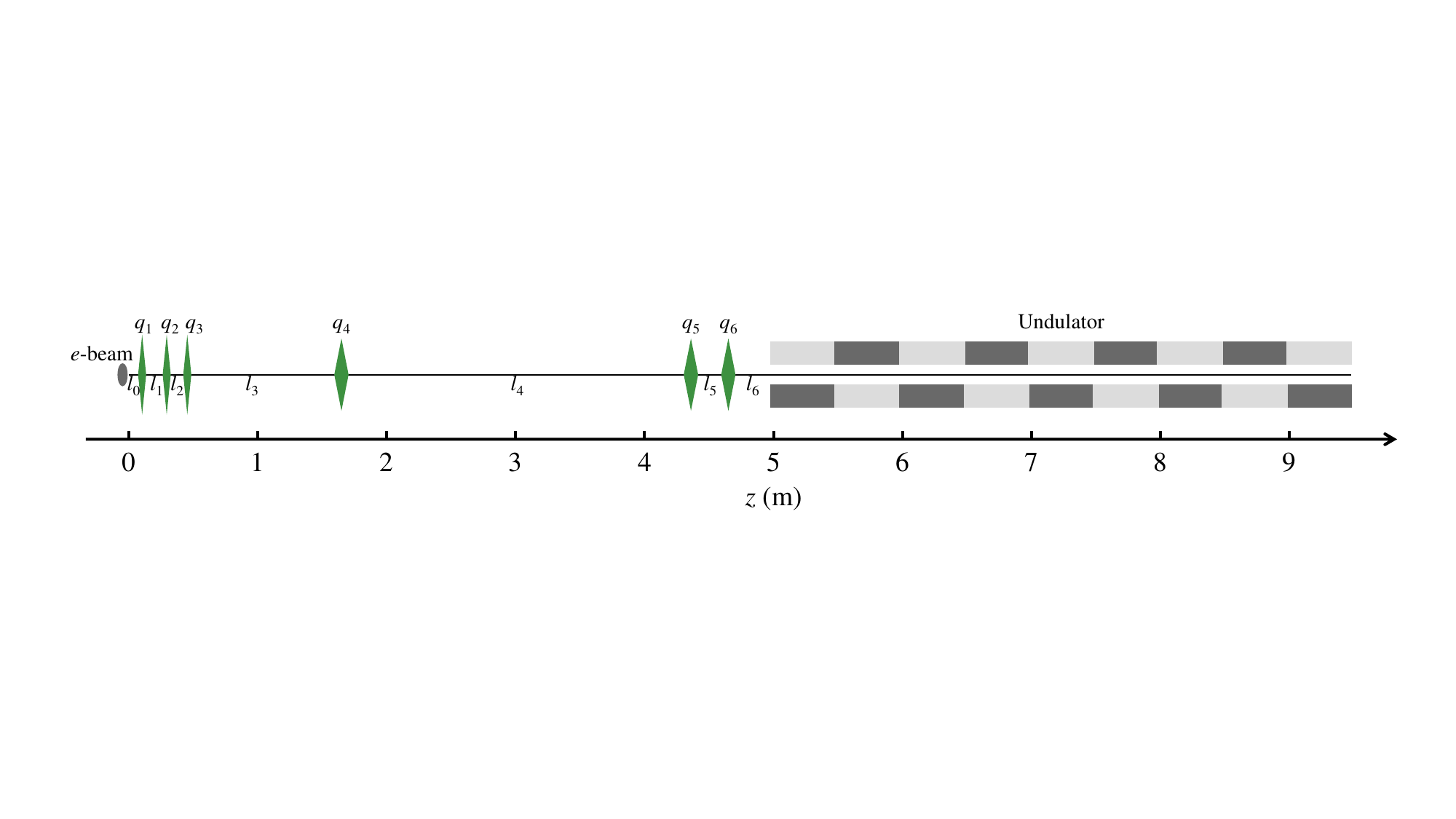}
    \caption{\label{fig:fig2}
    Schematic of the optimized beamline.
 }
\end{figure*}

Achieving reliable, high-gain LWFA-driven free-electron lasers (FELs) requires a beamline design robust enough to accommodate the shot-to-shot fluctuations of the \textit{e} beam caused by experimental jitters, as discussed in Fig.~\ref{fig:fig1}. The beamline consists of three 5-cm-long permanent-magnet quadrupoles, three 10-cm-long electromagnetic quadrupoles, and a 180-period planar undulator with a period length of 25 mm and an undulator parameter of $K_0=1.41$. The layout of the beamline is depicted in Fig.~\ref{fig:fig2}. 

\begin{table}
\caption{\label{tab:table1}
Parameter space used in $E_{min}$ optimization.
}
\begin{ruledtabular}
\begin{tabular}{cccc}
\textrm{Parameter}&
\textrm{Min. value}&
\textrm{Max. value}&
\textrm{Best value\footnote{Note $l_0-l_6$ in units of meters (m), $q_1-q_6$ in units of per square meter (m$^{-2}$).}}\\
\colrule
$l_0$ & 0.03 & 0.2 & 0.08\\
$l_1$ & 0.05 & 0.2 & 0.14\\
$l_2$ & 0.05 & 0.2 & 0.11\\
$l_3$ & 0.5 & 2.0 & 1.12\\
$l_4$ & 1.5 & 3.0 & 2.61\\
$l_5$ & 0.1 & 1.0 & 0.19\\
$l_6$ & 0.1 & 1.0 & 0.28\\
$q_1$ & -143 & -123 & -142\\
$q_2$ & 123 & 143 & 136\\
$q_3$ & -80 & -60 & -75\\
$q_4$ & -10 & 10 & -2.73\\
$q_5$ & 0.0 & 20 & 12.94\\
$q_6$ & -20 & 0.0 & -11.55\\
\end{tabular}
\end{ruledtabular}
\end{table}

We employ CMA-ES \cite{Hansen2016} implemented within the Optuna framework \cite{Akiba2019} as the primary optimization methodology. With the default setting, we adopt the theoretical optimum population size formulation $\lambda=4+3\ln n$ and standard adaptation rules for covariance matrix and step-size updates as established in the CMA-ES literature. The algorithm was initialized with a step size of 0.5 relative to normalized parameter bounds. The formulation of the objective function is crucial to the optimization process, as it serves as the mathematical embodiment of the target metric. Here, two distinct objective functions were implemented: the mean reciprocal value of the 3D power gain length $<L_G^{-1}>$ and the minimization of the simulated radiation energy $E_{min}$ obtained from the 13-electron-beam ensemble shown in Fig.~\ref{fig:fig1}. Both objective functions were strategically optimized within a systematic framework to achieve maximization of the desired performance metrics. The optimization parameters included the strength parameters of the quadruples and the distances between each element, arranged in sequence from upstream to downstream as follows: $l_0$, $q_1$, $l_1$, $q_2$, $l_2$, $q_3$, $l_3$, $q_4$, $l_4$, $q_5$, $l_5$, $q_6$, $l_6$. Here, $l_0$ represents the distance between the gas target and the first quadruple, $l_1-l_5$ denote the distances between adjacent quadruples, and $l_6$ corresponds to the distance between the last quadrupole and the undulator entrance. The variables $q_1-q_6$ represent the strength parameters of the quadruples. To accelerate the subsequent FEL energy optimization, the reciprocal of the gain length $<L_G^{-1}>$ was first maximized to identify a constrained set of viable beamline parameters. The detail of the $<L_G^{-1}>$ optimization were presented in Appendix~\ref{app:B}.  

\begin{figure}
    \centering
    \includegraphics[width=1\linewidth]{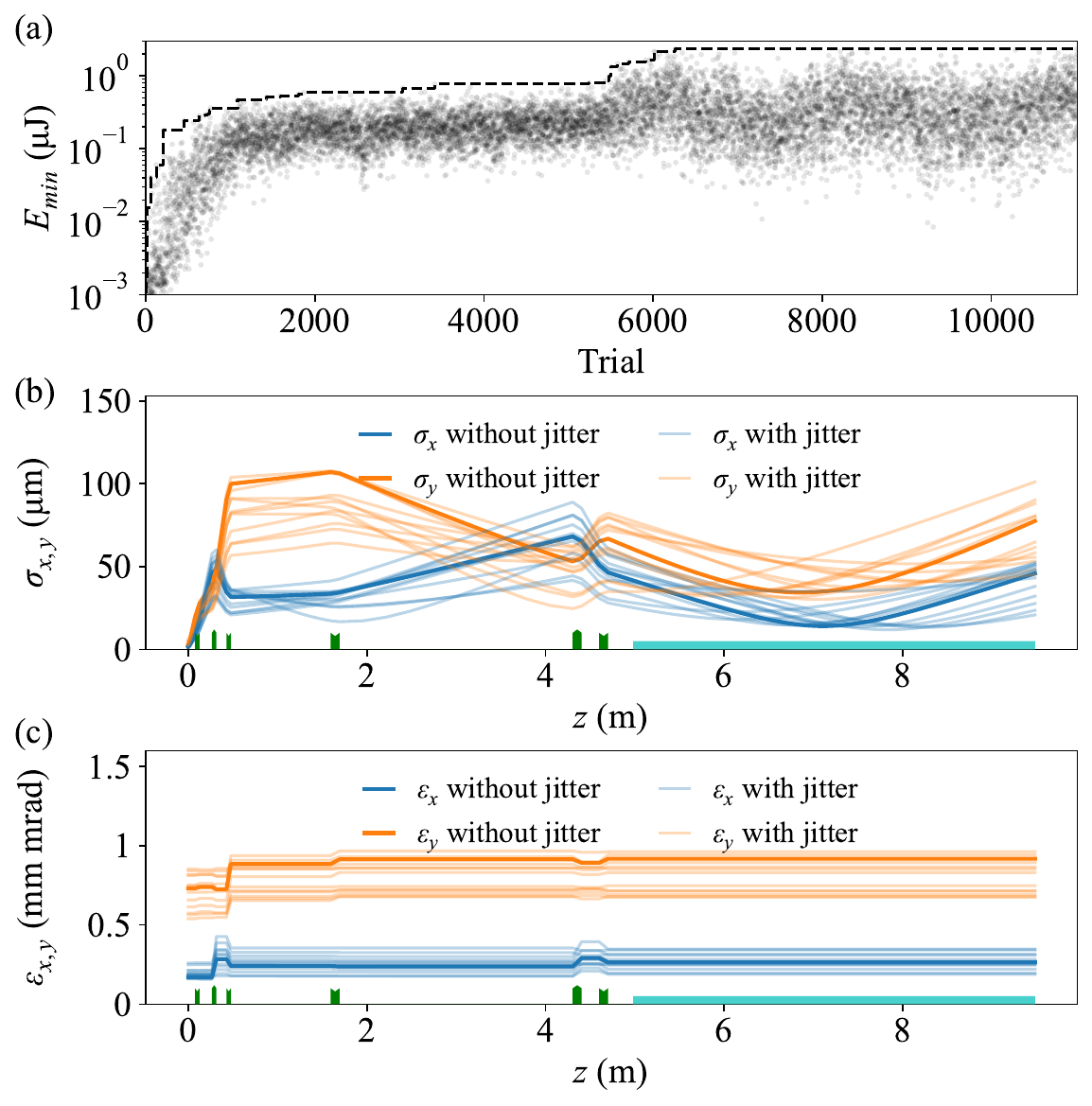}
    \caption{\label{fig:fig3}
    The optimization results of the $E_{min}$ for the 13-electron ensemble. (a) The simulated objective function $E_{min}$ (dots) with the cumulative best results (black dashed line). The y-axis is plotted on a logarithmic scale. (b)-(c) The evolution of the beam size (b) and the normalized projected emittance (c) along the beamline in horizontal (blue) and vertical (orange) directions. The tracking beamline was configured using the best parameters obtained from the optimization.
 }
\end{figure}

While minimizing the gain length $L_G$ establishes a beamline conducive to efficient FEL amplification, it can lead to distortion for FEL gain evaluation, as $L_G$ is evaluated from slice-averaged values of the beam. To address this, we shift the optimization objective to the minimum FEL energy $E_{min}$ expected within the 13-electron-beam ensemble show in Fig.~\ref{fig:fig1}. The results of the beam tracking obtained from ELEGANT code \cite{Borland2000} served as the input for subsequent FEL simulation performed with 3D and time-dependent code GENESIS \cite{Reiche1999}. The parameter space for this optimization, detailed in Table~\ref{tab:table1}, was systematically refined based on the results from $<L_G^{-1}>$ optimization.

\begin{figure*}
    \centering
    \includegraphics[width=1\linewidth]{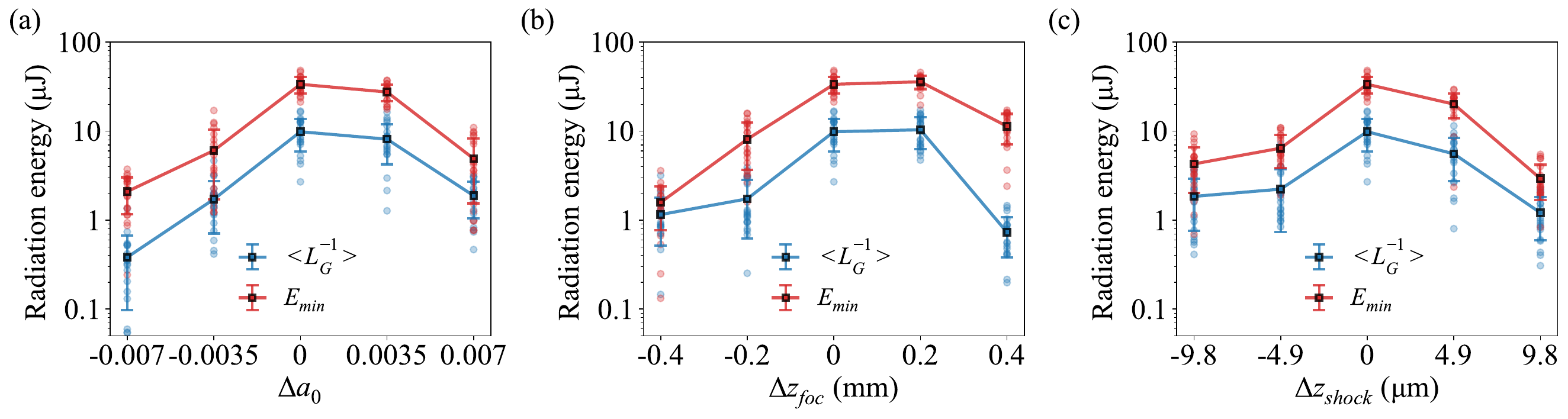}
    \caption{\label{fig:fig4}
    Radiation energy as a function of (a) normalized vector potential $\Delta a_0$, (b) laser focal position $\Delta {z_{foc}}$ and (c) the shock-front position $\Delta {z_{shock}}$. Mean values (black squares) and standard deviation (error bars) calculated from 20 independent simulations with varying random seeds, where the blue curve represents optimization using $<L_G^{-1}>$ as objective function and the red curve represents optimization using $E_{min}$ as the objective function. The y-axis is plotted on a logarithmic scale.
 }
\end{figure*}

\begin{figure}
    \centering
    \includegraphics[width=1\linewidth]{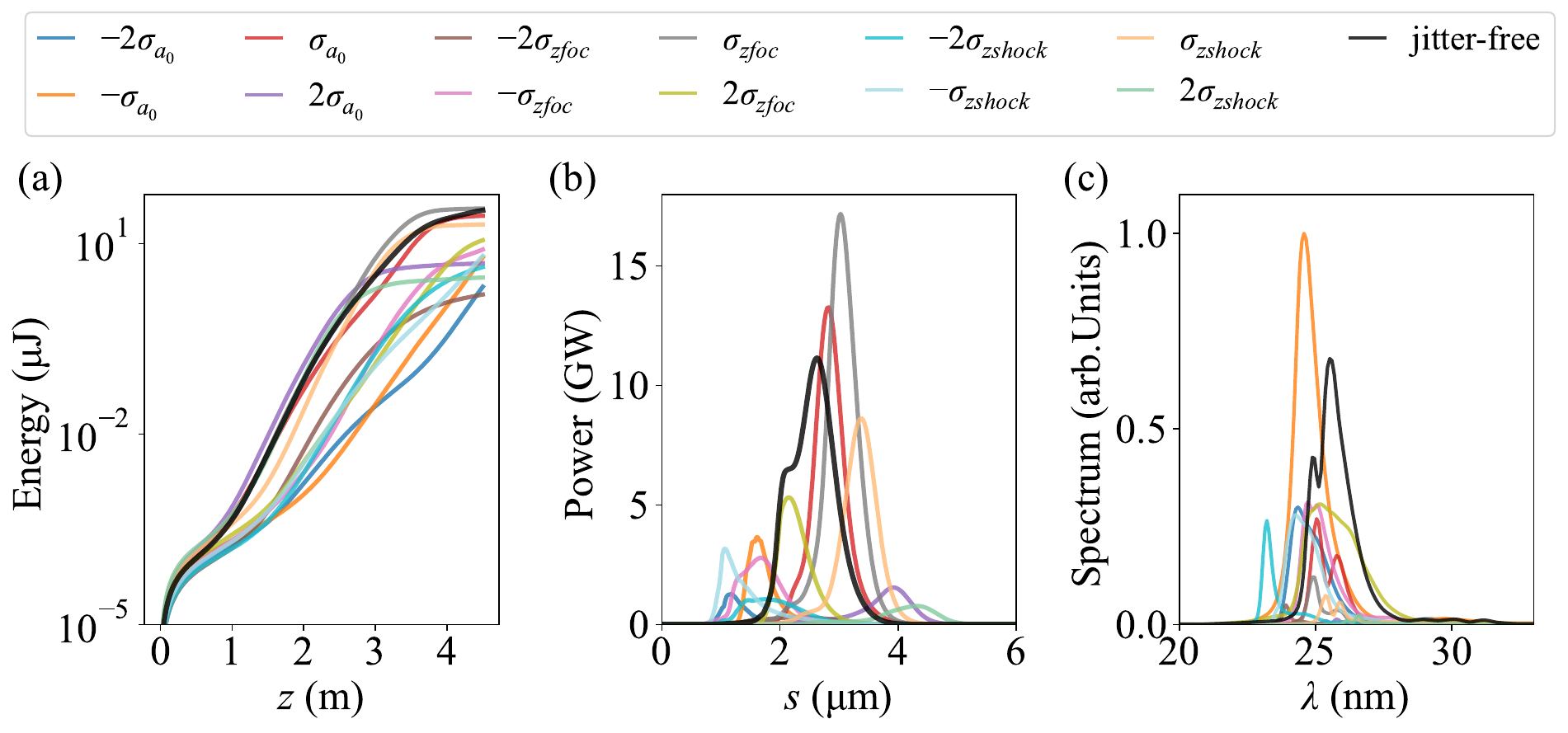}
    \caption{\label{fig:fig5}
    Simulation results for FELs with the 13-electron-beam ensemble. (a) Radiation energy along the undulator with the y-axis on a logarithmic scale. (b) Pulse duration and (c) spectrum of the generated radiation pulses at the exit of the undulator. Each line represents the average value of 20 distinct simulations with random seeds.
 }
\end{figure}

Figure~\ref{fig:fig3}(a) illustrates the evolution of the objective function $E_{min}$ as a function of iterations during the optimization. It is noted that the radiation energy for each evaluation was derived in a single simulation run using a fixed random seed for speed up. The optimal solution achieves at iteration 11,020, with the corresponding optimal parameters detailed in Table~\ref{tab:table1}. Under these optimized conditions, the radiation energy consistently exceeds 1 \textmu J for all electron beams in the ensemble, indicating robust operation of the LWFA-driven FELs in the saturated or high-gain regime(as shown in Fig.~\ref{fig:fig5}(a)) and effectively accommodating shot-to-shot fluctuations. The corresponding evolution of the beam sizes and transverse emittance along the beamline is presented in Figs.~\ref{fig:fig3}(b) and ~\ref{fig:fig3}(c), respectively. Figures~\ref{fig:fig4}(a)-~\ref{fig:fig4}(c) provide the comparison of the two optimization frameworks targeting $<L_G^{-1}>$ and $E_{min}$ under their respective optimized conditions. It is noted that the FEL energy was simulated with distinct random seeds for 20 runs in each situation in Fig.~\ref{fig:fig4}. While both frameworks exhibit similar trends in radiation energy, the $E_{min}$ optimized framework demonstrates superior performance with higher radiation energy. This enhancement can be attributed to two aspects, the fitting distortion of Eq.~(\ref{eq:2}) for the short-pulse situation induced by the slippage effects \cite{Pellegrini2016}, and the potential deviations introduced by using slice-averaged parameters for power gain length calculations. Through numerical simulations, the slice beam properties can be optimized to simultaneously achieve localized small averaged beam size, low energy spread, and high beam current, where the lasing occurs. Such optimization ensures more efficient electron-radiation interaction and facilitates energy extraction efficiency from the \textit{e} beam.

Figure~\ref{fig:fig5} presents the properties of the FEL radiation under the optimal operation targeting $E_{min}$. The evolution of the radiation along the undulator for the 13-electron-beam ensemble is shown in Fig.~\ref{fig:fig5}(a). A small subset of the beams enters the exponential-gain regime at about 1 m indide the undulator, and the majority begin exponential growth beyond 1.5 m and approach saturation or near-saturation by the undulator exit. Figure~\ref{fig:fig5}(b) illustrates the radiation power, highlighting localized lasing enabled by the carefully designed beamline. Through energy spread minimization and emittance matching, the beam parameters locally improve the phase-space characteristics, thereby enhancing the gain dynamics. Notably, even the minimum radiation power remains at the sub-gigawatt level, demonstrating strong tolerance to system fluctuations. The corresponding radiation wavelength fluctuates with an RMS value of 0.8 nm, as shown in Fig.~\ref{fig:fig5}(c) and Table~\ref{tab:table2}. This variation originates from energy fluctuations of the accelerated \textit{e}-beam, and thus the resonant wavelength. Detailed FEL simulation results for all 13 beam ensembles are summarized in Table~\ref{tab:table2}. It should be noted that the results presented in Fig.~\ref{fig:fig5} and Table~\ref{tab:table2} correspond to average values obtained from 20 independent simulations, each initialized with a random seed.

\begin{table}
\caption{\label{tab:table2}
Simulation results for FELs with the 13-electron-beam ensemble.
}
\begin{ruledtabular}
\begin{tabular}{ccccc}
\textrm{Parameter jitter}&
\textrm{Energy (\textmu J)}&
\textrm{Peak power (GW)}&
\textrm{Pulse duration (fs)\footnote{Note that the pulse duration is defined as the full width at half maximum (FWHM).}}&
\textrm{Wavelength (nm)}\\
\colrule
$-2\sigma_{a_0}$ & 2.09 & 1.26 & 1.40 & 24.35\\
$-\sigma_{a_0}$ & 6.06 & 3.65 & 1.45 & 24.58\\
$\sigma_{a_0}$ & 27.51 & 13.27 & 1.75 & 25.04\\
$2\sigma_{a_0}$ & 4.91 & 1.54 & 2.45 & 25.77\\
$-2\sigma_{zfoc}$ & 1.58 & 0.31 & 5.16 & 23.91\\
$-\sigma_{zfoc}$ & 8.10 & 2.79 & 2.85 & 24.73\\
$\sigma_{zfoc}$ & 35.84 & 17.18 & 1.78 & 24.96\\
$2\sigma_{zfoc}$ & 11.35 & 5.32 & 1.90 & 25.12\\
$-2\sigma_{zshock}$ & 4.28 & 1.05 & 3.87 & 23.22\\
$-\sigma_{zshock}$ & 6.44 & 3.17 & 1.48 & 24.28\\
$\sigma_{zshock}$ & 20.10 & 8.63 & 2.03 & 25.36\\
$2\sigma_{zshock}$ & 2.93 & 0.77 & 3.22 & 26.54\\
jitter-free & 33.61 & 11.15 & 3.19 & 25.52\\
\end{tabular}
\end{ruledtabular}
\end{table}

\section{Discussion and Conclusions}
\begin{figure}
    \centering
    \includegraphics[width=1\linewidth]{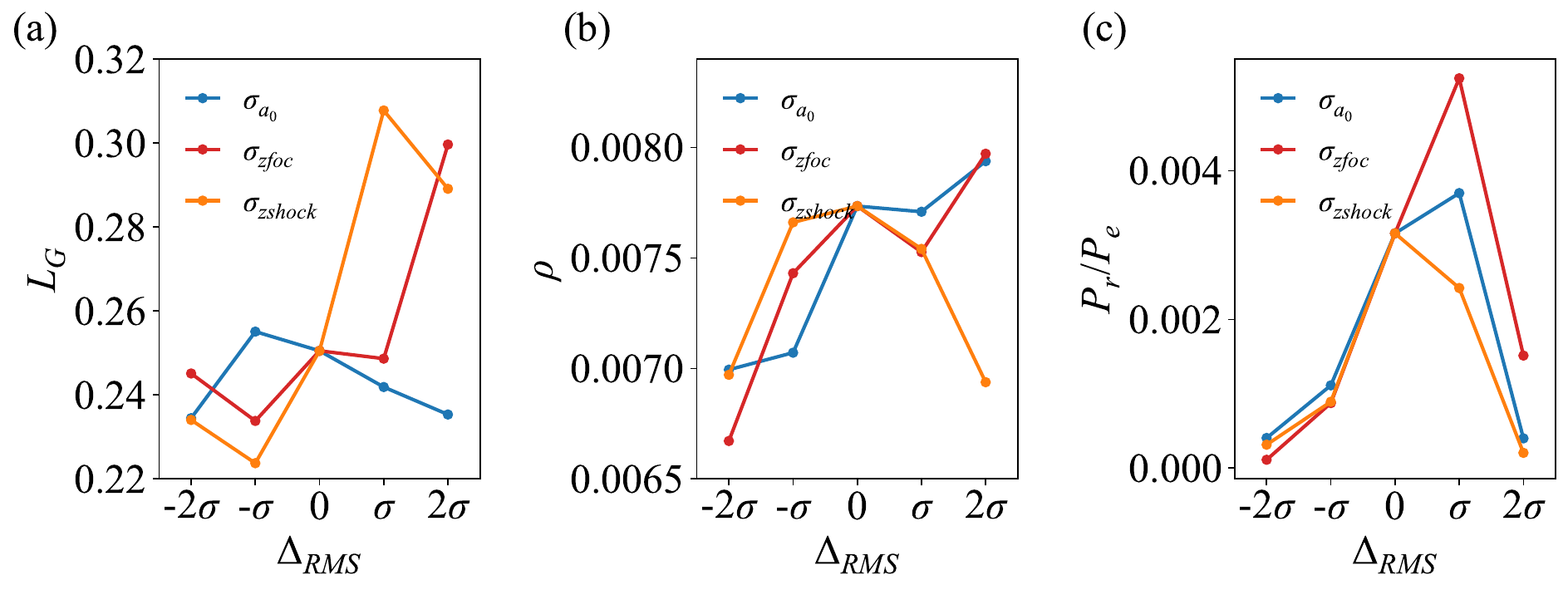}
    \caption{\label{fig:fig6}
    Parameters of the optimized electron beam: (a)  power gain length, (b) Pierce parameter, and (c) energy extraction efficiency. The blue, red and orange curves represent the laser energy variations, laser focal position displacement and shock front position instability, respectively. Horizontal axes denotes $\pm2$ folds RMS range of these three parameter jitters.
 }
\end{figure}

Figures.~\ref{fig:fig6}(a) and ~\ref{fig:fig6}(b) show the 3D gain length and the Pierce parameter under the optimized condition, and the corresponding calculation is detailed in Appendix B. Figure~\ref{fig:fig6}(c) presents the FEL energy-extraction efficiency, defined as the ratio of the peak radiation power $P_r$ to the peak electron-beam power $P_e=I\cdot E/e$, where $I$ is the peak current, $E$ is the beam energy, and $e$ is the elementary charge. The saturation length in a SASE FEL, defined as the undulator length at which the radiation power saturates, scales with the power gain length $L_G$. The scaling factor exhibits a logarithmic dependence on the number of electrons within one coherence time and typically lies within a narrow range of 18 to 20 \cite{Huang2007}. When accounting for specific parameter jitters in Fig.~\ref{fig:fig6}(a), the predicted saturation length exceeds 4.5 m. This theoretical prediction aligns with the results presented in Fig.~\ref{fig:fig5}(a), where the saturation is not reached within the considered undulator length for some cases. Extending the undulator would allow the radiation to progress from exponential gain regime to full saturation, thereby fulfilling its potential for higher energy output.

\begin{figure}
    \centering
    \includegraphics[width=1\linewidth]{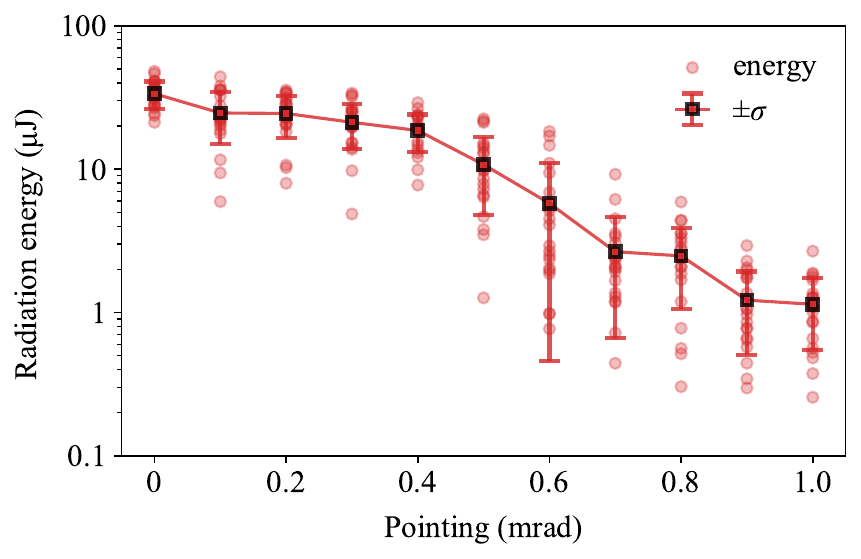}
    \caption{\label{fig:fig7}
    Radiation energy as a function of beam pointing jitter. Mean values (black squares) and standard deviation (red error bars) derived from 20 independent simulations (red dots) with random seeds. The y-axis is plotted on a logarithmic scale.
 }
\end{figure}

Beam pointing stability is another critical issue that must be addressed to ensure the robust operation of LWFA-driven FELs. To quantitatively evaluate the pointing instability, artificial angular kicks were introduced at the beamline entrance using \textit{e} beams simulated under jitter-free conditions ($\Delta_{RMS}=0$). Angular deviations in both transverse planes ($x^\prime:0.1-1.0$ mrad, $y^\prime:0.1-1.0$ mrad) were systematically introduced. Figure~\ref{fig:fig7} quantifies the resulting radiation energy degradation caused by this angular perturbation, revealing a monotonic decline in radiation energy from 34 \textmu J to 1 \textmu J as the pointing deviation increases to 1 mrad. Notably, the average radiation energy over 20 random seeds remains above 1 \textmu J even with 1 mrad deviation, indicating the proposed scheme's remarkable tolerance to angular deviations.

In conclusion, we present a systematic design and optimization of compact LWFA-driven FELs with enhanced robustness and reliability. The feasibility of the proposed scheme is demonstrated through comprehensive start-to-end simulations, which consistently yield FEL energies exceeding 1 \textmu J, even when accounting for laser and plasma instabilities. These simulations explicitly incorporate inherent parameter fluctuations, including laser energy jitter, focal position displacement induced by wavefront distortion, and shock front position instability in the plasma. The optimized configuration exhibits remarkable tolerance to these critical parameter fluctuations while maintaining optimal matching between the LWFA \textit{e} beams and the radiation field. These results establish crucial technical foundations for developing table-top FELs with unprecedented compactness, particularly highlighting robustness of the system against parameter jitter characteristic of current experimental configurations of LWFAs.

\begin{acknowledgments}
This work was supported by the National Natural Science Foundation of China (Grant Nos. 12388102), the Strategic Priority Research Program of the Chinese Academy of Sciences (Grant No. XDB0890200), the National Natural Science Foundation of China (12225411, 12474349, 12174410 and 12434012), CAS Project for Young Scientists in Basic Research (Grant No. YSBR060), CAS Youth Innovation Promotion Association (No. 2022242), and the New Cornerstone Science Foundation through the XPLORER PRIZE.
\end{acknowledgments}

\section*{Data availability}
The data that support the findings of this article are openly available \cite{DataAvailability}.

\bibliography{ref}

\appendix
\section{\label{app:A}Measurement of shock front}
\begin{figure}
    \centering
    \includegraphics[width=1\linewidth]{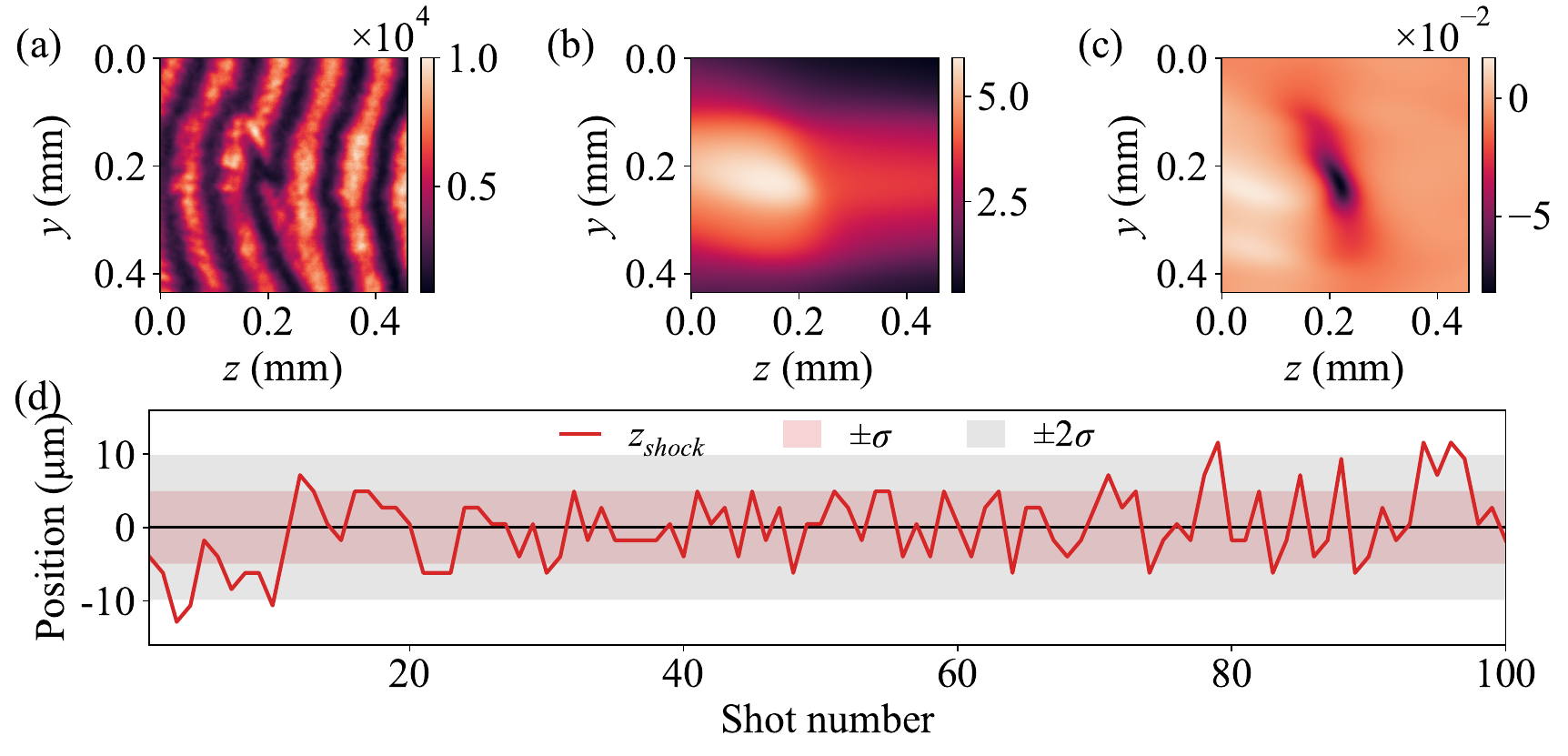}
    \caption{\label{fig:fig8}
    Properties of the measured shock front. (a) Measured interferometric fringe patterns and (b)-(c) the corresponding distributions of phase (b) and its first-order derivative (c). (d) Statics on the shock front jitter over consecutive 100 shots based on the first derivative of phase. The red and gray shaded region denotes the RMS and $\pm2$ times RMS range, respectively. 
 }
\end{figure}

The shock front is essential for the injection process in LWFA and thus significantly impacts the beam quality. As previously discussed, the position jitter with an RMS value of 4.9 \textmu m was used in our simulations, which was experimentally measured. Such a density profile can be achieved via the interaction of supersonically expanding helium gas with baffle. To capture the shock front properties, a probe laser passed through the plasma region and imaged using a 4f imaging system before entering the interferometer. The resulting interference patterns were captured using a charged-coupled device (CCD), with the shock front region depicted in Fig.~\ref{fig:fig8}(a). The corresponding phase distribution and the first-order derivative of the fringe phase are illustrated in Figs.~\ref{fig:fig8}(b) and ~\ref{fig:fig8}(c), respectively. To characterize the position jitter of the shock front, we performed continuous acquisition of 100 experimental shots, as shown in Fig.~\ref{fig:fig8}(d). An RMS value of $\sigma_{zshock}=4.9$ \textmu m was obtained. The statistical parameter is represented in the red shaded region in Fig.~\ref{fig:fig8}(d), providing quantitative analysis of the shock wave stability. 

\section{\label{app:B}Gain length optimization}
The 3D power gain length $L_G$ is a fundamental figure of merit for characterizing FEL performance, defined by $L_{G}=L_{G0}(1+\Lambda)$, where $L_{G0}=\lambda_u/(4\sqrt{3}\pi\rho)$ is the 1D power gain length, $\lambda_u$ is the period length of the undulator, $\rho$ is the dimensionless Pierce parameter defined by: 
\begin{eqnarray}
\rho=\frac{1}{\gamma_0}\left(\frac{1}{16}\frac{I_{p}}{I_A}\frac{K_0^2[JJ]^2}{K_u^2\sigma_x\sigma_y}\right)^{\frac{1}{3}}\;.
\end{eqnarray}
and $\Lambda$ is the degradation factor of gain length given by the fitting formula \cite{Huang2007}:
\begin{eqnarray}
\label{eq:2}
\Lambda&=&0.45\eta_d^{0.57}+0.55\eta_{\epsilon}^{1.6}+3\eta_{\gamma}^{2}+0.35\eta_{\epsilon}^{2.9}\eta_{\gamma}^{2.4}\\
&&+51\eta_d^{0.95}\eta_{\gamma}^{3}\nonumber+5.4\eta_d^{0.7}\eta_{\epsilon}^{1.9}+1140\eta_d^{2.2}\eta_{\epsilon}^{2.9}\eta_{\gamma}^{3.2}\;
\end{eqnarray}
where $\eta_d$, $\eta_{\epsilon}$ and $\eta_{\gamma}$ are three scaled parameters quantifies the diffraction, angular spread and energy spread effects, respectively, and can be written as: 
\begin{eqnarray}
\eta_d&=&\frac{L_{G0}}{2k_r\sigma_x\sigma_y}\;,
\\
\eta_{\epsilon}&=&\frac{2L_{G0}k_r}{\sqrt{\beta_x\beta_y}}\sqrt{\epsilon_x\epsilon_y}\;,
\\
\eta_{\gamma}&=&\frac{4\pi L_{G0}}{\lambda_u}\sigma_{\eta}\;.
\end{eqnarray}
where $\gamma_0$ is the electron energy in units of the rest energy, $I_p$ is the peak current, $I_A=17$ kA is the Alfv\'{e}n current, $K_0$ is the undulator parameter, $[JJ]=[J_0(\xi)-J_1(\xi)]$ with $\xi=K_0^2/(4+2K_0^2)$ is the Bessel function factor for a planar undulator, $K_u=2\pi/\lambda_u$, $k_r=2\pi/\lambda_r$, $\lambda_r=\lambda_u(1+K_0^2/2)/(2\gamma_0^2)$ is radiation wavelength, $\sigma_x$ and $\sigma_y$ is the average beam size in the undulator, $\beta_x$ and $\beta_y$ is the average Twiss parameter in the undulator, $\epsilon_x$ and $\epsilon_y$ is the average emittance in the undulator, and $\sigma_{\eta}$ is the relative RMS slice energy spread. The parameters $\gamma_0$, $I_p$, and $\sigma_{\eta}$ were obtained from FBPIC numerical simulations, whereas the transverse beam characteristics $\sigma_x$, $\sigma_y$, $\beta_x$, $\beta_y$, $\epsilon_x$, and $\epsilon_y$ were derived from particle tracking simulations using the ELEGANT code \cite{Borland2000}.

In the optimization framework targeting $<L_G^{-1}>$, we conducted parallel particle tracking simulations utilizing the 13-electron-beam ensemble illustrated in Fig.~\ref{fig:fig1}. The resulting transverse beam characteristics were subsequently employed to calculate $<L_G^{-1}>$, which serves as the primary quantitative metric for objective function evaluation. The parameter ranges for the optimization are presented in Table~\ref{tab:table3}. 
\begin{table}[h]
\caption{\label{tab:table3}Parameter space used in $<L_G^{-1}>$} optimization.
\begin{ruledtabular}
\begin{tabular}{cccc}
\textrm{Parameter}&
\textrm{Min. value}&
\textrm{Max. value}&
\textrm{Best value\footnote{Note $l_0-l_6$ in units of meters (m), $q_1-q_6$ in units of per square meter (m$^{-2}$).}}\\
\colrule
$l_0$ & 0.03 & 1.0 & 0.08\\
$l_1$ & 0.03 & 1.0 & 0.11\\
$l_2$ & 0.03 & 1.0 & 0.16\\
$l_3$ & 0.3 & 3.0 & 1.17\\
$l_4$ & 0.3 & 3.0 & 2.98\\
$l_5$ & 0.3 & 3.0 & 0.39\\
$l_6$ & 0.3 & 3.0 & 0.33\\
$q_1$ & -143 & 143 & -139 \\
$q_2$ & -143 & 143 & 141 \\
$q_3$ & -143 & 143 & -69 \\
$q_4$ & -45 & 45 & -0.34 \\
$q_5$ & -45 & 45 & 10.51 \\
$q_6$ & -45 & 45 & -10.82 \\
\end{tabular}
\end{ruledtabular}
\end{table}
\begin{figure}
    \centering
    \includegraphics[width=1\linewidth]{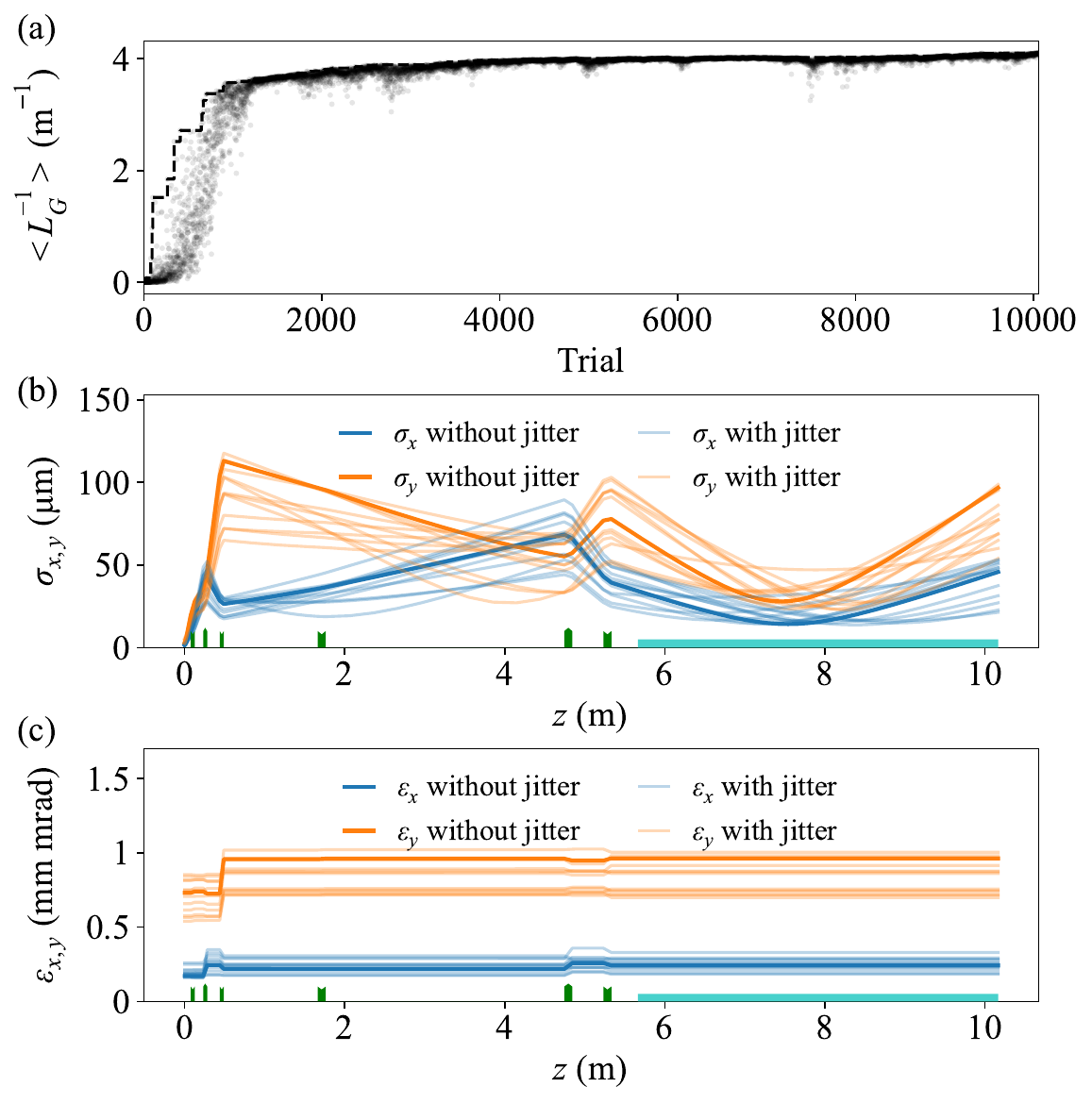}
    \caption{\label{fig:fig9}
    The optimization results of the $<L_G^{-1}>$. (a) The simulated objective function $<L_G^{-1}>$ (dots) with the cumulative best results (black dashed line). The y-axis is plotted on a logarithmic scale. (b)-(c) The evolution of the beam size (b) and the normalized projected emittance (c) along the beamline in horizontal (blue) and vertical (orange) directions. The tracking beamline was configured using the best parameters obtained from the optimization.
 }
\end{figure}

Figure~\ref{fig:fig9}(a) illustrates the evolution of the objective function $<L_G^{-1}>$ throughout the BO process, with the cumulative optimal results represented by the black dashed line. The optimization commenced with random sampling within the predefined parameter space detailed in Table~\ref{tab:table3}. A progressive convergence of the objective function emerges after approximately 1200 iterations, with the optimal solution achieved at iteration 10,071, corresponding to an average 3D power gain length of approximately 0.24 m. The resulting beam sizes and transverse emittance for the 13-electron-beam ensemble under the optimized conditions are presented in Figs.~\ref{fig:fig9}(b) and ~\ref{fig:fig9}(c), respectively. It should be noted that slice-averaged energy spread of the \textit{e} beam is employed when calculating the power gain length. The radiation gain predominantly originates from specific slices with advanced beam qualities, while other slices contribute minimally to the overall gain. The optimization based on slice-averaged values may cause potential deviations, considering the different properties over slices of the \textit{e} beam from LWFAs. To address this limitation, the optimization using the radiation energy as an alternative objective function is implemented. 

\end{document}